**High Velocity Dust Collisions: Forming Planetesimals in a Fragmentation Cascade with Final Accretion**

Jens Teiser and Gerhard Wurm

Institut für Planetologie
Wilhelm-Klemm-Str. 10
D-48149 Münster, Germany
E-mail: j.teiser@uni-muenster.de
Tel.: (+49) 251 8339047
Fax.: (+49) 251 8336301

**Abstract**

In laboratory experiments we determine the mass gain and loss in central collisions between cm to dm-size $SiO_2$ dust targets and sub-mm to cm-size $SiO_2$ dust projectiles of varying mass, size, shape, and at different collision velocities up to ~56.5 m s$^{-1}$. Dust projectiles much larger than 1 mm lead to a small amount of erosion of the target but decimetre targets do not break up. Collisions produce ejecta which are smaller than the incoming projectile. Projectiles smaller than 1 mm are accreted by a target even at the highest collision velocities. This implies that net accretion of decimetre and larger bodies is possible. Independent of the original size of a projectile considered, after several collisions all fragments will be of sub-mm size which might then be (re)-accreted in the next collision with a larger body. The experimental data suggest that collisional growth through fragmentation and reaccretion is a viable mechanism to form planetesimals.

## 1. Introduction

The formation of planetesimals, km-size objects which further evolve to planets, is still an unsolved problem. While it is clear that somehow dust particles embedded in protoplanetary discs have to gather themselves to form planetesimals, the task is not a trivial one and all models suggested have shortcomings so far.

One group of models invokes gravitational attraction of solids. A basic problem is to achieve the high densities necessary for gravitational instability to form planetesimals. The original idea of a dust laden sub layer in the disk by Safronov (1969) and Goldreich & Ward (1973) was challenged as it turned out that turbulence generated by shear would prevent densities high enough to be reached (Weidenschilling, Donn & Meakin 1989). Current work to find solutions to overcome this problem is ongoing (Youdin & Shu 2003; Yamoto & Sekiya 2004). Recently, Johansen, Klahr & Henning (2006) showed that turbulence might be helpful under certain conditions as it can lead to clumping of larger "boulder" size bodies and further gravitational binding if those boulders would already exist in a large amount. As yet unaccounted for in this model is the collisional behaviour of the "boulders" and the initial conditions, i.e. a large number of decimetre bodies would have to have formed by other means first.

The necessary formation of decimetre objects naturally leads to the second group of models, where planetesimal formation is a growth process, mostly built on mutual collisions and sticking. Several authors have attempted to solve the coagulation (growth) problem by different methods (Weidenschilling & Cuzzi 1993; Weidenschilling 1997; Ormel, Spaans & Tielens 2007; Brauer, Dullemond & Henning 2008; Johansen et al. 2008). Different levels of turbulence are assumed within the disc, which, together with different drift motions, generate collisions. In the literature more or less plausible outcomes of collisions are assumed, i.e. perfect sticking or fragmentation e.g. producing power law size distributions of solids or bimodal size distributions. Depending on these assumptions, growth is easy, is too slow or gets stalled at certain sizes. Collisional behaviour between dusty bodies is therefore a central part of all models and decides if planetesimal formation is allowed in a certain model or prohibited.

However, sticking or fragmentation is not a completely free model parameter. It is determined by the collisional physics. It was shown experimentally and theoretically over the last years that the growth of cm-size aggregates from dust is almost inevitable and fast, essentially independent of the disk model (Dominik & Tielens 1997; Wurm & Blum 1998; Kempf, Pfalzner & Henning 1999; Blum & Wurm 2000; Blum et al. 2000; Poppe, Blum & Henning 2000a; Colwell 2003; Paszun & Dominik 2006; Wada et al. 2007; Blum & Wurm 2008, Suyama, Wada & Tanaka, 2008). The growth process might be aided by more sticky frosty or organic material (Kouchi et al. 2002; Bridges et al. 1996; Supulver et al. 1997). Once larger than cm, fragmentation occurs and has to be considered for further evolution (Blum & Wurm 2000). Collisional charging and electrostatic reaccretion have been proposed for growth to proceed beyond the fragmentation limit (Poppe, Blum & Henning 2000b; Blum 2004; Blum & Wurm 2008). Also reaccretion by gas flow has been proposed for this stage (Wurm, Blum & Colwell 2001a,b). In earlier experiments and models Wurm et al. (2001a, 2001b), Wurm, Paraskov & Krauss (2004) and Sekiya & Takeda (2003) showed that an initially eroding impact can still lead to net growth as the gas flow will transport small slow fragments back to the target. The final size of the objects capable of growing by gas aided reaccretion is depending on the internal structure of the growing objects and is still debated (Wurm et al. 2004; Sekiya & Takeda 2005). Most likely it is possible to generate at least decimetre size objects by these processes in the inner disk.

The next step in a simple hit-and-stick scenario would be collisions between the decimetre size objects and smaller particles at velocities exceeding 10 m/s (Weidenschilling & Cuzzi 1993; Sekiya & Takeda 2003). In a laminar disk collisions would be restricted to a maximum of about 60 m s$^{-1}$ between different size particles throughout the whole disc (e.g. Weidenschilling & Cuzzi 1993, Weidenschilling 1977), once the larger body is several decimetre in size, depending in detail on the disk model and distance to the star. If the disk profile would be rather flat, somewhat reduced collision velocities would be possible. In regions and times where dust densities would be comparable to gas densities, collision velocities might also be reduced, at least for a certain time, as the solids would take the gas along (Weidenschilling 1997). On the other hand, in strongly turbulent models significant collision velocities might also be induced by turbulence and include collisions between equal size larger objects (Johansen et al. 2008). Whatever the details, collisions of several tens of m/s are likely and the question remains if collisional growth can proceed through such a violent phase at all.

The high collision velocities have frequently been and are still regarded as an obstacle for planetesimal formation as it seemed unlikely that growth can be the outcome of such collisions. Indeed fragmentation has already been observed for fractal dust aggregates at about 1 m s$^{-1}$ (Blum & Wurm 2000). More compact bodies do not fragment that easy but bounce off each other up to several m s$^{-1}$ which has been observed in a number of experiments by now as well (Blum & Muench 1993; Wurm, Paraskov & Krauss 2005). Bouncing is also observed in the experiments described here, though we will not analyse this further. We note that the fact that compact dust aggregates bounce off each other at low speed – and do not stick – might be crucial for the formation of planetesimals as outlined later.

Nevertheless, growth is possible at high collision velocities. The highest collision velocities at which growth between dust aggregates had been observed so far was in experiments by Wurm et al. (2005). Between 13 m s$^{-1}$ and 25 m s$^{-1}$ compact dust aggregates of a few mm in size (projectiles) collide with a compact dust target of several cm in size. The projectiles on average leave 50 per cent of their mass sticking to the target. These results were verified in the drop tower not to be influenced by gravity (Paraskov, Wurm & Krauss 2007). While 25 m s$^{-1}$ is fast, it does not unambiguously allow growth of planetesimals in protoplanetary disks under typical conditions, i.e. at still higher speeds and for larger projectiles. With this in mind, we carried out experiments at higher collision velocities and for different projectile parameters to continue the former work by Wurm et al. (2005), where the highest velocity was limited by the capabilities of the projectile launcher.

2.    **Experimental Setup**

In Fig. 1 the principle setup can be seen. The experiments are carried out in a vacuum chamber to avoid effects of gas drag during the impact and on the ejected particles.

Pressure during impacts was below 10$^{-1}$ mbar, typically a few times 10$^{-2}$ mbar. To avoid target explosions by out gassing, evacuation was set to a moderate rate and took several hours. We tested if collisions at still lower pressure would prevent target destruction by evacuating for 15 hours to below 10$^{-3}$ mbar but no difference in the outcome of collisions was visible.

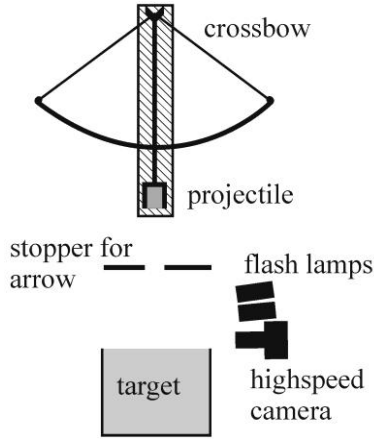

**Fig. 1.** Experimental setup. All components seen except the flash lamps and camera are within a vacuum chamber. The target is placed on foam or in a string cradle to prevent vibrations of the impact of the arrow onto the stopper to couple into the target. The bottom of the projectile holder is covered with aluminium foil which bends upon impact at the stopper and releases the projectile.

The experiment chamber is equipped with a crossbow launcher. Arrow tips are metal plates to hold dust aggregates. The plates are stopped about 25 cm above the target as they impact a mechanical structure with an inner clearing. The dust then moves through this central clearing due to inertia. A dust projectile of cylindrical shape launched that way can be seen in Fig. 2.

The image is part of a high speed movie (500 full frames per second) which captures each impact. Two flash lamps are used for illumination, each runs with 500 Hz but both are shifted by 1 ms with respect to each other. Therefore, each image is a double exposure (see Fig. 2). From this timing and the distance between the two projectile images, its velocity is determined. The projectile then impacts a target. Targets have been placed in string cradles or placed on foam to avoid a significant coupling of impact vibrations of the launcher stop into the target.

As target we used cylindrical containers filled with dust material compressed manually to a volume filling of ~ 33 per cent. The dust consists of $SiO_2$ grains (quartz) with a size range of 0.1 – 10 μm (80 per cent between 1 and 5 μm). This is in a range of particle sizes typically considered for preplanetesimal dust. Depending on the impact velocity and the projectile size, different target sizes are used to prevent the far field damage described below.

The projectiles consist of the same compressed dust material. Table 1 gives an overview over all performed experiments discussed in this work.

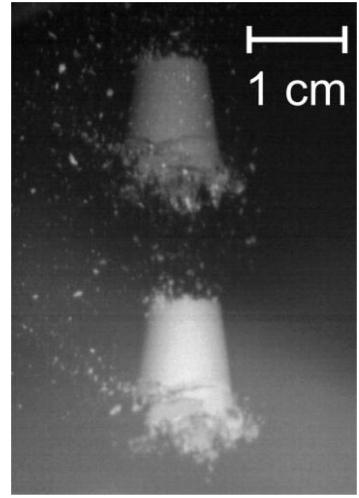

**Fig. 2.** Projectile imaged by two subsequent flashes with 1 ms time delay. The travel distance then gives the velocity of 24.7 m s$^{-1}$. The original slightly conical shape of the projectile holder is well maintained.

As the exact composition of preplanetesimal dust agglomerates is not known we assume that the dust used in these experiments is a good analogue material for dust in protoplanetary discs as the mechanical properties of dust aggregates are determined by the grain size distribution and hardly depend on the chemical composition of the monomeres (Blum et al. 2006, Langkowski, Teiser, Blum 2008). Due to mutual collisions fractal aggregates which grow in the first place get more and more compact with increasing size (Blum & Wurm 2008). With increasing aggregate sizes also the velocities grow larger (Weidenschilling & Cuzzi 1993) so relative velocities studied in this work should occur only for highly compressed bodies as described above.

For large cylindrical shaped projectiles dust is compressed into the projectile mount and fixed by a thin aluminium foil to prevent it slipping out of its mount. The mm-size projectiles are irregular shaped fragments of a larger piece of compressed dust. Flat projectiles are made by pressing and spreading a pile of dust on the surface of a flat metal plate used as projectile mount until a certain thickness is reached. The projectile is then fixed by several stripes of aluminium foil.

To distinguish the projectile material from the target material after the impact some dust used for the projectiles was coloured grey using black ink. Larger pieces of compressed dust consisting of coloured dust are slightly more brittle qualitatively compared to the original white dust projectiles of similar size but in the frame of this paper this has no significant influence on the impact results as seen in the following sections.

The target is weighed before the experiment after it spent a day at low, constant humidity (22 per cent). It is weighed after the impact after another day at constant humidity. Any free fragments on the surface are then removed either by tilting the target 90° or by careful vacuum cleaning.

Fast fragments leave the target in a flat angle and are not able to hit the target again but stick to the walls of the experiment chamber after the experiment. As detailed in earlier studies by Wurm et al. (2005), slow fragments returning to the target after an ejection do not make firm contact, i.e. do not stick if they are slower than 6.7 m s$^{-1}$. Fragments which hit the target again are observed to be much slower in the experiments reported here. In fact they are typically seen to bounce off several times before they eventually get to rest on the target. They are easily removed from the surface in contrast to material which sticks on the surface due to the impact. The targets are photographed afterwards.

Depending on the total target mass the accuracy of the mass determination varies from ± 5mg for lower mass targets (m < 600 g) and ± 20 mg for more massive targets due to humidity variations. Due to tilting/vacuuming the accurancy of the mass determination would be additional 10 per cent of the projectile's mass (Wurm et al. 2005).

From the mass determination we calculate accretion efficiencies when possible defined as

$$e_{ac} = \frac{m_{Ta} - m_{Tb}}{m_P},$$

(eq. 1)

where $m_T$ is the target mass after and before the impact, $m_P$ is the projectile mass. Values of $e_{ac}$ have to be below 1. The extreme $e_{ac}=1$ would be a complete accretion of the projectile. Negative values imply erosion of the target. In this work we use the accretion efficiency only to divide between growth and erosion. Within the limits of the mass determination this is accurate for the massive projectiles while we currently cannot quantify the mass balance for an individual 1mm projectile. For our collision experiments with an array of small projectiles (marked as multiple in table 1) the mass variation of the target is in the range of the sum of the projectiles' mass. Therefore, these experiments (see table 1) are only discussed qualitatively.

According to the observed outcome of the experiments, we distinguish 3 different aspects of the impact which seem to be characteristic, (1) the fate of the projectile material, (2) the near field effects of the impact onto the target in the vicinity of the impact site which extents 2 to 3 projectile sizes, and (3) the far field effect of the impact on target material further away.

## 3. Projectile Fragmentation and Accretion

In former experiments by Wurm et al. (2005) it was found that above 13 m s$^{-1}$ but below 25 m s$^{-1}$ impact speed a projectile of a few mm in size fragmented when impacting the target. About half of the mass of the original projectile was found to be left in firm contact with the target material afterwards as a cone shaped structure. Here we find that somewhat more massive and larger projectiles do create a crater within the target but depending on the impact parameters a central cone shaped part often remained as well. To distinguish between effects on the projectile and the target we carried out impact experiments onto a solid steel plate and in other experiments we used coloured dust as projectile material.

| No | v (m/s) | projectile mass (mg) | Thickness (mm) | type | target | accretion efficiency |
|---|---|---|---|---|---|---|
| 1 | 52.5 | 631 | 8.6 | Normal | Dust | -3.0 |
| 2 | 47.9 | 612 | 8.5 | Normal | Dust | -1.6 |
| 3 | 44.1 | 618 | 8.5 | Normal | Dust | -0.21 |
| 4 | 34.5 | 347 | 7.5 | Normal | Dust | -3.3 |
| 5 | 44.0 | 318 | 6.8 | Normal | Dust | -2.1 |
| 6 | 44.2 | 274 | 6.5 | Normal | Dust | -1.6 |
| 7 | 44.2 | 223 | 6.1 | Normal | Dust | -3.1 |
| 8 | 24.8 | 886 | 9.6 | Normal | Dust | -0.91 |
| 9 | 26.7 | 895 | 9.6 | Normal | Dust | -0.89 |
| 10 | 27 | 180 | 4 | Normal | Dust | -0.4 |
| 11 | 36.8 | 12 x 24 | 3 | Multiple | Dust | Negative |
| 12 | 32 | 12 x 10 | 2 | Multiple | Dust | Negative |
| 13 | 35 | - | 1 | Multiple | Dust | Positive |
| 14a | 53.5 | - | 0.5 | Multiple | Dust | Positive |
| 14b | 53.5 | - | 1 | Multiple | Dust | Negative |
| 15 | 44 | - | 1 | Disrupted | Dust | Positive |
| 16 | 30.2 | 286 | 1.0 | Flat | Dust | 0.052 |
| 17 | 45.0 | 729 | 1.5 | Flat | Dust | -0.056 |
| 18 | 39.5 | 1623 | 2.0 | Flat | Dust | -0.5 |
| 19 | 41.5 | 1338 | 1.5 | Flat | Dust | -0.058 |
| 20 | 46.8 | 1074 | 1.0 | Flat | Dust | -0.61 |
| 21 | 40 | 1465 | 1.5 | Flat | Dust | -1.41 |
| 22 | 43 | - | 0.5 | Cloud | Dust | Positive |
| 23 | 56.5 | - | 0.5 | Cloud | Dust | Positive |

| 24 | 22.6 | 325 | 6.9 | Normal | Plate | 0.28 |
| 25 | 24.0 | 850 | 9.5 | Normal | Plate | 0.34 |
| 26 | 27.1 | 885 | 9.6 | Normal | Plate | 0.25 |
| 27 | 24.7 | 849 | 9.5 | Normal | Plate | 0.20 |
| 28 | 47.4 | 1635 | 2.5 | Flat | Dust | Complete Destruct-tion. |

**Table 1**. Overview over the different experiments with different projectile and target properties performed in this work. Experiment 28 is shown in fig. 7. Otherwise experiments with far field damage (complete destruction) are not included.

Projectile parameters are mass, shape, porosity with only mass and shape being varied in the experiments. In experimental studies with non-cohesive materials impact energy often is the important parameter (e.g. Colwell et al. 2008). As shown below this is not true for experiments with cohesive dust. We therefore already note here that projectile thickness (extension of the projectile in direction of motion) turns out to be an important parameter for the outcome of a collision.

### 3.1 Steel plate impacts

A 1 cm projectile impacting a steel plate adds 20 to 34 per cent of the mass (experiments 24 to 27, see table 1). Figure 3 shows an example of a dust projectile impacting onto a metal plate (exp. 25). The cone like structure which was observed before, occurs here at a collision velocity of 24.0 m s$^{-1}$.

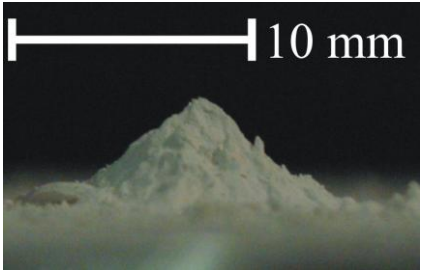

**Fig. 3.** Result of a cylindrical projectile of 9.5 mm height and 8.5 mm diameter impacting a steel plate (exp. 25).

In agreement to earlier studies the fragment distribution was extremeley flat in the impact on steel plates (Fig. 4).

In images captured when half of the projectile already had made contact with the steel plate, it is seen that the end of the projectile proceeds unchanged and is not influenced by the destructive processes ahead at the steel plate. Fig. 5 shows an image sequence of an impact (exp 25).

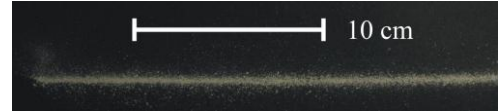

**Fig. 4**. Fragments as found on the wall of the experiment chamber for the impact onto a steel plate (exp.25). The line of dust traces the tangential to the steel plate. In agreement to earlier findings by Wurm et al. (2005) for impacts onto dust targets most dust is ejected within 1° of the tangential to the plate surface.

As the final height of the cone is reached, it can be seen that the projectile splits up, creating the final cone shape as material is ejected to the sides. In a simplistic view, it was speculated before by Wurm et al. (2005) that the material escapes upon contact at the front edge of the projectile but not in the centre part of the projectile and that the cone builds up as material from the rear of the projectile fills empty space at the edge. This is obviously not the case and the outer shells only split off at later times in the collision process.

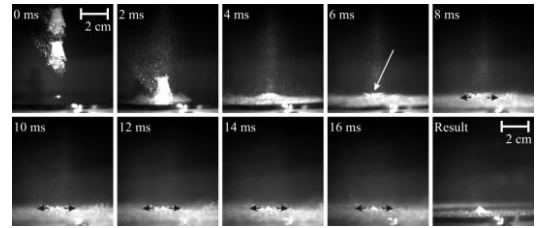

**Fig. 5.** Sequence of images showing the impact of a projectile onto a steel plate (exp. 25). The top of the projectile remains intact and cylinder like until the final height of the cone eventually forming is reached. The side parts of the cylinder only split off as a final process of the impact leaving a cone sticking behind. The dust cone forms in the first frames of this sequence and does not grow further but is hidden by the fragment cloud which is still vanishing after 16 ms.

In a second impact of a same size projectile, the projectile did hit exactly the same spot with the underlying dust cone (exp 26). This resulted in a somewhat higher cone and left 25 per cent of its mass sticking to mass already present. We measured a third projectile of a similar size impacting the steel plate at a different location (exp. 27), which left 20 per cent of its mass sticking.

Obviously projectiles of the given make-up always end up as cone like remnants which contain about a quarter of their original mass. This is comparable to the result by Wurm et al. (2005) where on average half of the projectile mass was found to stick. We have to note though that the impacts there were not directed onto a steel plate but onto a dust target. In those experiments also the projectile was not as well constrained in shape. What fraction of the mass difference is due to the target or projectile is currently unknown.

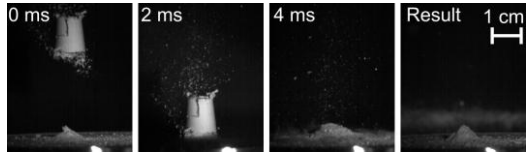

**Fig. 6.** Multiple collision (exp. 26 on top of the dust cone of exp. 25): A projectile hits the dust cone from a previous impact and results in a similar mass gain, essentially only increasing the size and mass of the existing dust cone.

However, the choice of the target material (steel plate or compact dust) does not seem to be of major importance for the initial interaction between projectile and target. If the target is already a compacted dust aggregate, the projectile dust particles will be decelerated as quickly as by a steel plate, resulting in a partial destruction of the projectile upon contact and formation of a cone in both cases.

### 3.2   Coloured projectiles

In Fig. 7 the result of a flat projectile (thickness 2.5 mm, diameter 30 mm) of coloured dust impacting a target of 10 x 2.3 cm in size can be seen (experiment 28, see table 1).

The target was not extended enough to prevent global damage (see section 4.2 below). However, the fragments of the target clearly show a layered structure. The dust fragments consist of a white original target material to which the grey projectile dust is firmly attached (fig. 7).

The fragmentation process does not separate target material from projectile dust already sticking to it. This implies that the compressive and tensile strength of the new aggregate consisting of projectile and target material are similar through the whole body.

Projectile dust did essentially not get transported over a larger scale. The target shows the grey colour in a confined region corresponding to the impact site. It is a general observation for the flat projectiles that the grey target area corresponds well with the original size of the projectile. Little material from the projectile itself is taking part in the ejection process.

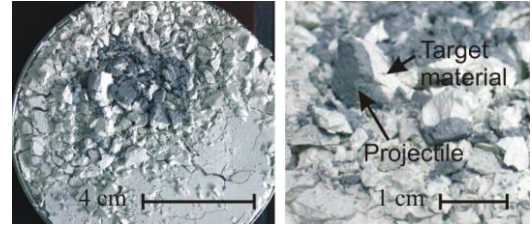

**Fig 7.** Fragments of a destroyed thin target and projectile in a high speed collision at 47.4 m s$^{-1}$ (experiment 28, see table 1). The grey material is the projectile material of a flat but massive projectile (diameter: 30 mm, thickness: 2.5 mm, mass: 1.635 g). On the right, projectile dust can be seen firmly attached to the target material in a layer consistent with the original projectile thickness. On the left it can be seen that the fragments with projectile matter attached are comparable to the original diameter of the flat projectile.

Taken all together, we conclude that in a first step part of the projectile sticks to the target while the propagation of damage within the target can be treated separately. If no target damage occurs the target will gain several tens of percent in mass depending on the projectile shape, i.e. a cylinder like projectile will lose more than half its mass while a flat projectile might more or less stick completely.

### 4.   Impact damage of the target

### 4.1   Near field target damage

In a qualitative picture, if the collision is energetic enough, damage to the target occurs. Fig. 8 shows three impacts that have different projectile parameters, namely a number of small (~3 mm) projectiles impacting at 36.8 m s$^{-1}$ forming trenches of various depth around the projectile cone (no. 11, see table 1), a flat projectile of 3 cm diameter and 2 mm thickness impacting at 39.5 m s$^{-1}$ (no. 18) and a massive cylindrical projectile at 34.5 m s$^{-1}$ forming a crater which includes the impact site as material loss (no. 4). The mass of the flat projectile (no. 18) is larger than the mass of cylindrical projectile (no. 4) but the mass loss in case (no. 4) is much more severe. The accretion efficiencies are -0.5, and -3.3 for (no. 18) and (no. 4) respectively. Obviously the projectile shape and especially the thickness play a significant role in the process of near field damage.

As discussed before, it is visible that the projectile adds a significant part of the mass to the target in case (no. 11) (cones) and (no. 18). Net mass gain occurs if the damage to the target is small, essentially if no damage occurs and not in the cases shown though the erosion is very minor. Impact (no. 4) was too energetic and the original cone was destroyed as well.

It is important to note that within the resolution of the high speed movies and in the covered parameter range it is found that ejecta of cylindrical projectiles or projectiles with similar extension in all three dimensions are always smaller than the original projectile. An example is seen in fig. 9.

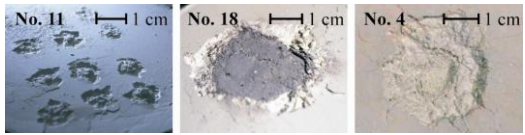

**Fig. 8.** Near field damage in different types of collisions: (no. 11) a spatially separated field of ~3 mm projectiles which collided at 36.8 m s$^{-1}$ shows trenches around central projectile cones. (no. 18) a flat projectile of 2.0 mm thickness and 30 mm diameter impacting at 39.5 m s$^{-1}$ shows target damage surrounding the projectile. (no. 4) a large 7.5 mm thick cylindrical projectile impact at 34.5 m s$^{-1}$ leaves a crater without central projectile cone.

This is also consistent with the profile through a crater which at maximum is half as deep as the projectile's original size. In the given parameter range the projected area of the craters reach maximum values of 2-3 projectile diameters.
This relation is independent of the projectile size. Similar extensions of impact craters are found in Colwell et al. (2008) though the target material is quite different and less cohesive. Mostly target material, not projectile material takes part in the slow ejection process if not the whole sticking projectile mass is removed in very energetic events. The size of the near field damage with respect to the projectile size also implies that ejecta have to be smaller than the original projectile. Exceptions are the flat projectiles where the projected area is much larger than the thickness and near field damage might be larger than the thickness. However, as already the fragments of such a collision are no longer very flat, this might be regarded as non-typical dust aggregates.

Wurm et al. (2005) give a size distribution of projectile fragments, where the fragment size is much smaller than the original projectile size. According to their work the maximum sizes in the fast ejecta were found to be only about 10-20 per cent of the projectile size. Otherwise the size distribution was found to be flat down to about 100 μm. If only the projectile parts are ejected, fragments have to be smaller than the projectile per definition, but we argue that this is also true for target ejecta within the given parameter space.

Further evidence for smaller fragments comes from the accretion efficiencies which vary between –0.2 and –3.3 at maximum for the large massive projectiles. Not much more mass than the original projectile mass is available. This mass has to be distributed among a large number of fragments. If one fragment was bigger than the projectile it should clearly be visible in a cloud of much smaller particles. In the movies of eroding collisions reported here we find only evidence for ejecta smaller than the projectile in near field damage.

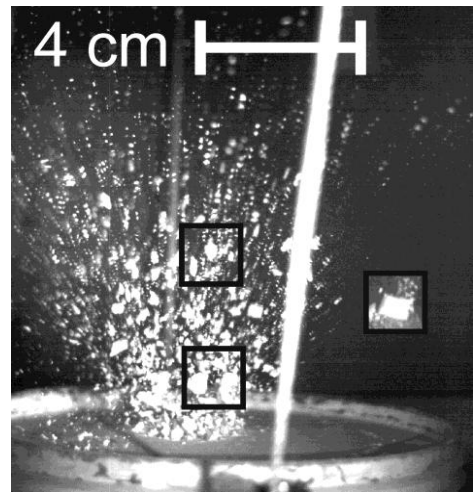

**Fig. 9.** Image of airborne ejecta after a destructive collision with a 6.1 mm thick cylindrical projectile (m = 0.223 g) at 44.2 m s$^{-1}$ (experiment 7, see table). The original projectile has been embedded in the image for size comparison. All visible ejecta are smaller than the projectile.

We are confident that we did not miss individual ejected fragments larger than the projectile but clearly note that, so far, this has to be regarded as a qualitative statement, as we did not systematically study the fragment distribution. Due to the initially dense field of slow fragments and the limit of resolution for smaller projectiles a dedicated setup might be needed to image and collect the fragments which was not the original focus of this work.

For the model discussed below it is crucial that on average, for many collisions, not more larger fragments are ejected in erosive collisions than the number of the initial projectiles or erosion could continue indefinitely and destroy large bodies. However, an occasionally large fragment will not be a problem for the model. A weak statement that ejecta are e.g. most of the time not larger than the projectile size is sufficient to allow larger bodies to grow. As fragments from one collision are the projectiles for the next collisions, a cascade of high speed collisions will provide ever smaller projectiles. If projectiles stick below a certain size, such a cascade will result in net growth eventually. Nevertheless, special emphasis has to be put on the fragment distribution in the future.

### 4.2 Far field target damage

In some experiments with the most massive projectiles the target was damaged far beyond the local impact site. Cracks run through the target and ejecta were observed disconnected from the impact site. In the most destructive events the target was only 2.3 cm thick and 10 cm in diameter had a mass of ~ 219.236 g while the projectile has itself 1.635 g (experiment 28, Fig. 7). Obviously the target was not thick enough to damp the energy which is released as elastic waves through the target upon impact. However, after we increased the size of the target to ~ 7 cm thickness and 17.5 cm diameter, such extreme damage was no longer seen for the same kind of projectiles. With a target mass being 1000 times larger than the projectile mass no far field damage will occur (all data points in figure 10 show only impacts with near field damage). The size of our target is then still smaller than the size of objects for which such high impact velocities as several tens of m s$^{-1}$ are expected in protoplanetary disks (Weidenschilling & Cuzzi 1993). We conclude that for objects larger than 10 cm in size only the near field damage is important and impacts are a local phenomenon on the surface for the given projectile parameter range. Cracks in the target, if they occur due to impacts, might influence further evolution of a target but this is beyond the analytical capabilities of the current experiments.

### 5. Erosion and Accretion

Fig. 10 shows a summary of the collisions between dust projectiles and dust targets which lead to net growth ($e_{ac} > 0$) or net erosion ($e_{ac} < 0$) of a target depending on the projectile velocity and thickness. Only collisions without far field damage are plotted because only those impacts are considered to be relevant for protoplanetary discs. Impact velocities are determined with an error of less than a few per cent. Collisions which lead to a mass gain of the target are marked by full circles. Mass loss corresponds to open symbols. The different projectile types are listed by letters.

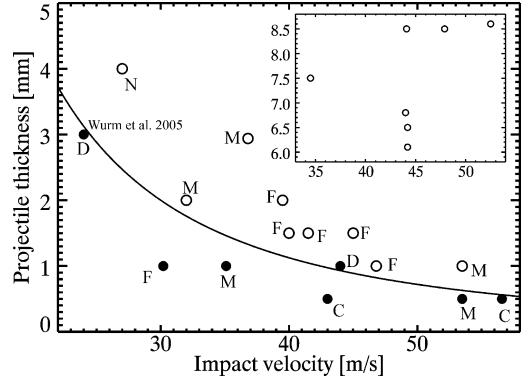

**Fig. 10.** Accretion versus erosion depending on impact velocity and projectile thickness. Open symbols are erosive impacts (negative accretion efficiency), closed circles mark positive accretion efficiencies. The main plot emphasizes the threshold between erosion and accretion. The inset in the upper right is a somewhat dispatched region of high velocity and large thickness impacts with cylindrical projectiles. The symbols mark different projectile types; N: normal cylindrical, M: multiple impacts of spatially well separated projectiles, F: flat projectiles, C: clouds of particles, D: disrupted projectiles (v = 24 m/s: result from Wurm et al. 2005). The solid line is a model for the threshold adapted to the data (see text for detail).

The inset in the top right corner has normal cylindrical projectiles (symbol N) which were large of about 1 cm in thickness as seen in fig. 2. These always lead to erosion of the target but for a target body of several decimetres in size an individual collision does not change the mass and size significantly. The largest mass losses are on the order of 3 times the projectile mass or volume, which is about 3 cm$^3$. To erode a 30 cm radius body by only the outer 1 cm would require about 1000 collisions with 1 cm projectiles. Therefore, this set of experiments shows that the original target is not significantly altered by such collisions if it would only interact with its own fragments.

As stated before, the ejecta of the corresponding near field damage are always smaller than the original projectile. Subsequent collisions will be with smaller projectiles, which will result in still smaller projectiles and so on. If we only assume the projectiles to be 2 mm smaller than the original projectile, about 5 collisions are needed to get all particles to be smaller than 1 mm, which is a maximum mass loss of the target of 3$^5$ cm$^3$ = 243 cm$^3$. As a large number of fragments will already be smaller than 1 mm after each collision, this is a worst case estimate, showing that a target is not significantly eroded even if it would interact with its own ejecta continuously. After this erosion all projectiles of the next impacts are sub-mm in size.

However, continious growth requires that the population of cm-size particles is not dominating. In this case the target bodies might not reaccrete very small particles produced by the grinding, but will be eroded by collisions with different cm-size particles.

The larger projectiles impacting onto dust targets lead to a mass loss, while similar impacts onto steel plates lead to mass gain (cone formation). Experiments with coloured dust showed that some projectile material always sticks to the target surface.

If some projectile material always sticks to the target, then the distinction between mass gain and mass loss of the target is determined by the target damage.

The experimental data show a transition between growth and erosion at smaller sizes. The size of 1 mm is roughly the threshold below which projectiles are (re)-accreted at high speed, as can be seen in Fig. 10. Our setup currently does not allow measuring the exact mass balance for individual collisions with small projectiles of 1 mm as they only carry 1 mg of mass or less. This is on the level of uncertainty due to the large dust mass of the target and humidity effects associated with its large surface area even if the target is kept at constant humidity. Different approaches were taken with our given setup to approach the small projectile size limit.

To be able to measure mass differences still, flat projectiles were used, marked by (F) in Fig. 10. The flat projectiles had only 1 mm to 2 mm thickness but the mass balance was well measurable due to the extension of the projectile which was on the order of 30 mm. This projectile shape is not assumed to be relevant in protoplanetary discs as these experiments were performed for a better understanding of the dynamic processes during the impacts. As detailed below we argue that the onset of near field damage is mostly determined by the thickness of a projectile.

A few experiments were carried out with up to 12 individual small projectiles which were launched in parallel but spatially well separated onto the same target. This allows to measure an average mass gain or loss and/or to give a qualitative comparison for a number of individual impacts. Such experiments are marked (M) for multiple projectiles. For projectile thicknesses of 1 mm also this method has its limit. In two such cases we judged qualitatively from the images if the accretion efficiency was positive or not (exp. 13 and 14). At the high impact velocity of 53.5 m/s (exp. 14), some projectiles in an M-type experiment clearly created a crater with more volume than the original projectile. At the same time other impact sites of the same experiment showed almost no crater, where we assume growth has taken place. Supposedly, these mark the transition region between erosion and accretion. We splitted this experiment in two data points. As there is no measurement of a mass balance for this data point, we have to note that there is some ambiguity in these values.

We also studied disrupted projectiles (D), where the projectile breaks up into a large number of smaller fragments during launch. The disrupted projectiles at 24 m s$^{-1}$ represent 3 impacts from Wurm et al. (2005), where we estimated the typical fragment size from the airborne projectile images. In these experiments projectiles and targets were produced by Wurm et al. (2005) with the same technique and the same materials as in this study. Due to a different accelerator (spring instead of a crossbow) the projectiles had not a cylindrical shape (as the N-type experiments) but broke to smaller fragments (D-type). The given size in the diagram refers to the size of the typical projectile fragments in these experiments. As the results of Wurm et al. (2005) are gained from three impact experiments showing the same characteristic the outcome of this study has to be considered when applying the model plotted in the diagram and described below to the experimental data.

In one D-type experiment (exp. 15) we had a projectile of initially 3 mm height fragmenting during launch into smaller fragments. From the high speed movie we estimate the characteristic fragment size to be 1 mm. The net effect of this impact was positive accretion. Therefore, growth of the dominating particle size of about 1 mm is possible at 44 m s$^{-1}$.

In 2 other experiments we produced extended and mass loaded clouds of particles (C) by launching a projectile through a 2 mm mesh at 43 m s$^{-1}$ and 56.5 m s$^{-1}$ (exp. 22 and 23). These clouds could easily be measured to lead to significant mass gain of the target. This was also visible as large parts of the surface were completely covered with firmly sticking coloured dust which was used in these experiments. This partially complete coverage with projectile dust and the large extension of the cloud in width implies that interaction between ejecta of the foremost cloud projectiles with later projectiles of the cloud might take place. However, this will only be important for the dense region of the cloud. At the edge of the particle cloud projectiles should interact with the target as individual particles. In fact, very little damage could be noted in the transition region between optically thick coloured dust and the original white target material in the cloud cases under a stereo microscope. We noted many cone-like features of projectile dust without any target damage where the sticking projectile had 1 mm or less in diameter. From these images and the image of some distinct particles within the diffuse cloud impacting, we estimate that 0.5 mm is a typical size for the projectiles within the particle cloud. Very few craters of 2 mm or larger were seen, which we attribute to impact of projectile parts within the cloud somewhat larger than 1 mm. On average, the mass gain and the absence of near field damage for small projectiles indicates that sub-mm aggregates accrete to a larger body at the given velocities. However, further experiments with dedicated setups for individual small projectiles are needed to confirm this and are planned for the near future.

## 5.1 A model for the threshold thickness/velocity relation

The following simple model gives our current expectations on the threshold between erosion and accretion with respect to projectile thickness (size) and collision velocity.

We start by noting that we measured the sound speed for the compact targets which we used to be 90 m s$^{-1}$ (± 10 m s$^{-1}$). This was done by a runtime measurement giving an acoustical pulse on one side of the dust sample and using a force sensor to detect the pulse signal on the opposite side measuring the runtime with a digital oscilloscope.

Projectile impacts are therefore not supersonic. We assume that our compact targets are elastic and show an abrupt transition between elastic compression and material failure which will result in the formation of target crater and ejecta once a maximum stress $E_{max}$ has been applied to the target by the projectile which is decelerated. We further assume that the projectile of mass m acts like one solid body. The basic equations in the elastic regime then are the equation of motion and Hooke's law.

$$F = ma = m\ddot{x} \qquad \text{(eq. 2)}$$

$$F/A = Dx \qquad \text{(eq. 3)}$$

where A is the cross section of the projectile hitting the target, D is an elasticity constant which has the unit of a pressure per length, and x is the depth of compression. The solution to eq. 2 and eq. 3 is a standard harmonic oscillation of the projectile position with time t after first contact with the target.

$$x = v_0 \sqrt{c} \sin(\sqrt{1/c}\, t) \qquad \text{(eq. 4)}$$

$$\ddot{x} = -v_0 \sqrt{1/c} \sin(\sqrt{1/c}\, t) \qquad \text{(eq. 5)}$$

$$c := m/AD \qquad \text{(eq. 6)}$$

It is $v_0$ the projectile velocity upon impact. If the maximum local stress applied by the projectile is less than the strength needed to fracture the target ($E_{max}$), the projectile will rebound if it stays intact. This is e.g. observed for compact projectiles below a certain velocity of several m s$^{-1}$, where they indeed stay intact, do not fracture largely and rebound (Wurm et al. 2005). If the projectile fractures largely, its kinetic energy is dissipated efficiently and it can stick to the target.

However, if the maximum stress $E_{max}$ of the target is reached the target will fracture as well. Mass loss will occur. We take this condition as threshold between accretion and erosion. It should be noted that $E_{max}$ in our calculations might not equal the compressive strength of the material as we do not have a unidirectional compression only. The target material outside of the impact site provides some support for sidewards motion. The projectile does not have any support. Therefore, the projectile, though equally compact, fragments at much lower velocities. At the threshold failure will occur at maximum compression and we get as condition

$$\frac{E_{max}}{\sqrt{\rho D}} = v_0 \sqrt{d} \qquad or \qquad \frac{E_{max}}{\sqrt{D}} = v_0 \sqrt{\frac{m}{A}}, \quad \text{(eq. 7)}$$

where we assume a constant density ρ for the projectile and d being the thickness of the projectile. Eq. 7 suggests that the threshold between accretion and mass loss due to target failure in the near field of the impact site is given by the impact velocity times the square root of the projectile thickness being constant for a given material.

In the ideal case of a largely extended projectile like the flat projectiles, the inner parts of the compressed target area are only compressed along the impact direction and for symmetry reasons no shear between neighbouring spots is present. On the other side, at the edge of the impacting projectile shear along the impact direction occurs within the target. Therefore, damage is first done surrounding the projectile area, while no damage occurs below the projectile as observed for flat projectiles.

This is a very simple model, but calibrated by experiments it might provide some predictions for a limited parameter space and for not too much extrapolation. More accurate modelling is needed but until that is given eq. 7 might be used as simple analytical equation to discriminate between growth and erosion. A line in agreement to the experimental data is

$$d = \frac{a}{v_0^2} \quad \text{with a = 1800 m}^3\text{s}^{-2} \quad \text{(eq. 8)}$$

This is plotted in fig. 10. In experiments with granular targets it is found that the ejection process depends on the kinetic energy of the projectile (Colwell et al. 2008). In granular materials the attractive forces between single grains are small in comparison to gravity which for dust materials is completely different. Dust materials are dominated by the attractive forces between the single dust grains so the mechanical properties neither are comparable to the mechanical properties of solid bodies nor to those of granular materials. We note that for the near field damage in dust impacts the mass or impact energy is not the basic parameter. Massive but thin projectiles lead to growth while cylinders of comparable (or even less) mass do not (see table 1, impacts no. 10 and 16). The shape of the projectile obviously is of some importance.

## 6.    Planetesimal growth

We found in the experiments that at collision velocities of several tens of m s$^{-1}$ up to velocities expected in protoplanetary disks of 50 m s$^{-1}$ to 60 m s$^{-1}$, 1 mm is a typical threshold size between erosion and accretion. Larger projectiles lead to erosion but smaller projectiles add mass to a target upon which they impact. One might ask if the destructive nature of large projectile impacts implies that an existing large body would eventually be destroyed. In view of what is known about collisions from experiments so far, this is not necessarily the case. On the contrary, a larger existing body is able to continue to grow as outlined here. We suggest a model which is based on the sticking properties and collision velocities of the different dust aggregates participating in the collisional process.

With the initially slow collision velocities of dust particles it is known that particles about decimetre in size can grow (Blum & Wurm 2008). This process sets off as fractal growth process but with increasing size aggregates become more and more compact. Even in mutual collisions with impact velocities of less than 1 m/s the collision partners are compacted significantly (Blum & Wurm 2008) and fractal aggregates are of minor relevance for further growth processes.

Once the aggregates reach decimetre size compaction has created bodies which are as compact as the targets we used here. The experiments carried out here did not deal with this initial growth phase of decimetre bodies but consider how further growth might proceed. Three different categories of collisions have to be considered then.

(a)    collisions between large (>decimetre) bodies of equal size
(b)    collisions between large and small bodies (>decimetre target, <cm projectile)
(c)    collisions between small particles (both <cm)

Collisions (a) do especially occur in turbulent disks and can lead to collisions of meter-size bodies of 10 m s$^{-1}$ or more (Weidenschilling & Cuzzi 1993). We assume that they are very destructive resulting in a significant part of much smaller bodies. However, they are at least occasionally needed to provide smaller particles needed as reservoir to grow larger bodies. If continuous growth of the existing larger bodies by small bodies is guaranteed then these collisions are not obstacles to planetesimal formation. Eventually, the number of large objects will decrease and catastrophic destructions will get less frequent.

At the same time collisions of type (b) will – as net effect – lead to the growth of the existing larger bodies. This is suggested by the experiments reported above. Collision velocities in this category are at several tens of m s$^{-1}$. The cm-size aggregates will certainly erode a larger body first but, as argued above, it will not change its size significantly. On the other side, the ejected aggregates will be smaller than the impacting aggregate. We did not study all different parameters important for collisions yet (e.g. impact parameter, dust particle size, target roughness …) but this statement is based on the experiments carried out so far. We assume that this holds generally. After a few collisions the fragments are then small enough (<1mm) that they are (re)-accreted in the next set of collisions.

The fragments produced are very likely not individual dust grains or very small aggregates to a large extent as it gets increasingly difficult to produce very small aggregates due to the large number of contacts which have to be broken between the grains. In fact Wurm et al. (2005) found that the fragment distribution after a high speed collision is not a power law with increasing particle numbers towards smaller aggregates but is flat. While not specified in Wurm et al. (2005) their particle collectors were not indicating a large number of unresolved individual dust particles. This and the measured size distribution show that very small aggregates do not constitute a large fraction of the mass. This is important as collisions with very small dust aggregates consisting of up to a few individual micron-sized grains might be erosive (R. Schräpler & J. Blum, personal communication). In total, the collisions of type (b) are driving the growth process. The `larger` cm-aggregates are grinded to sizes between 0.1 mm to 1 mm and are then accreted by larger bodies, which act as catalyser for the grinding first.

Category (c) collisions are also important. The very initial formation of large bodies starts by sticking of small dust aggregates in gentle collisions. Such gentle collisions of fractions of m s$^{-1}$ to several m s$^{-1}$ occur in this category of small particles (Weidenschilling & Cuzzi 1993). If the small aggregates ejected in collisions could grow again by mutual collisions they would eventually provide larger cm-size projectiles again, which would continue to erode a large decimeter body. This would counteract the grinding. The number of grinding collisions would not be limited and net growth would not be guaranteed. However, the particles ejected from collisions with compact targets are compact aggregates. In contrast to fractal dust aggregates compact (even highly porous) aggregates do not stick at the given velocities. They only rebound or continue to fragment slightly (Blum & Muench 1993; Heisselmann & Blum, personal communication; Wurm et al. 2005). Mutual collisions of small aggregates will therefore not change the size distribution but be neutral. It should be noted that it is an important finding of earlier works that slow collisions of compact aggregates do not lead to growth. This is beneficial as the size distribution of small particles can only be changed then by collisions with the large decimetre (or larger) bodies. Particles of centimetre size and smaller (projectiles) can only get smaller. The model is visualized in Fig. 11.

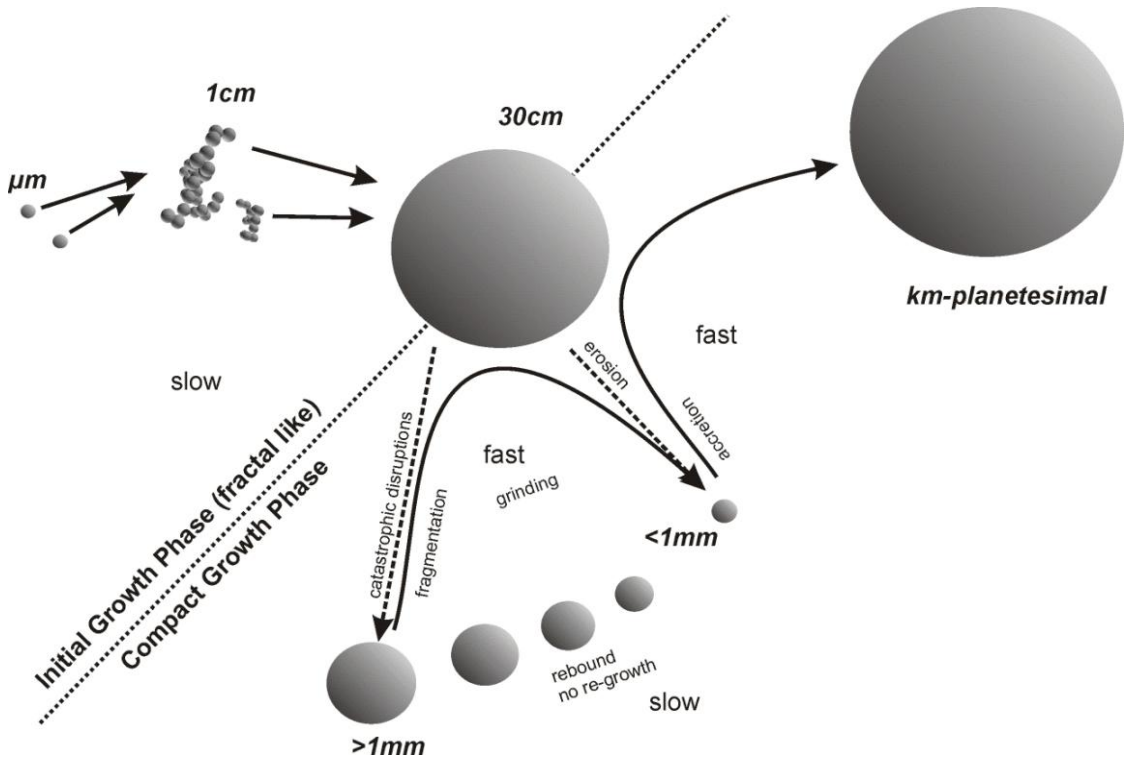

**Fig. 11.** A model for planetesimal formation. Ejecta in fast collisions only get smaller and do not grow again in mutual collisions. Eventually ejecta get small enough to be (re-) accreted by a large object.

Johansen et al. (2008) considered a similar situation theoretically in a coagulation / fragmentation calculation. They assumed that all small particles which interact with a large body are accreted instantly. While our experiments show that this is not exactly true, the final result after several steps of a fragmentation cascade would be the same. Therefore, our model essentially shares all possibilities and shortcomings as discussed in Johansen et al. (2008).

The model does not solve the problem of drift time scales, which prevails for all coagulation models so far as particles drift too rapidly inwards (Weidenschilling 1977; Brauer et al. 2008; Johansen et al. 2008). The radial drift is an omnipresent problem in almost all models (except in fast gravitational instability models). It has also been shown by Johansen et al. (2008) that the small particles after a collision would diffuse again vertically in strong turbulence which reduces the collision rate and growth rate. However, this might only prolong the growth time scale if the drift could be stopped. Certainly more moderate conditions like a dead zone with no or weak shear induced turbulence would be beneficial for times of planetesimal formation. If radial drift is not present or of minor importance, e.g. in high dust density regions, then planetesimal formation through collisional growth seems feasible. Our experiments show that a growth model is possible in principle even including high speeds, if not partially requiring them. Growth in collisions is not ruled out in the first place as sometimes assumed for high impact energies. Still further experiments and modelling are required to detail the accretion efficiency especially of fast sub-mm particles and to specify the fragment size distribution for larger fast projectiles in more detail. Certainly other parameters as impact angles should be varied as well to get a more complete picture of collisions eventually.

We started our targets as compact targets with only 67 per cent porosity. We regard this as appropriate as a successive bombardment with small particles at intermediate velocity will build such compact targets which is subject to our current research and beyond this paper. It has to be noted that the bulk of our particles are 1-5μm in size. While 1 μm is often considered as initial building block size this is much larger than interstellar particle sizes. Smaller dust particles might just as well have constituted the original dust population in protoplanetary disks. This should help increasing sticking considerably (Blum & Wurm 2008).

## 7. Conclusions

We found in experiments that a collision of sub-cm projectiles with a compact target at collision velocities of several tens of m s$^{-1}$ has 3 distinct features.
- the addition of projectile mass to the target
- the removal of target material in the vicinity of the impact site
- global damage to the target, if the target is small

For targets of several decimetres the latter point is not significant.

Particles smaller than 1mm can add mass to a preplanetesimal body at up to 56.5 ms$^{-1}$. Extrapolation and qualitative indications from the experiments suggest that probably sub-mm aggregates can add mass to a target beyond 60 m s$^{-1}$. Larger projectiles will result in damage at this velocity. The experiments further show that
- the fragments produced during the impact are typically much smaller than the projectile
- the fragments are compact dust aggregates and the results of published experiments suggest that such aggregates do not stick to other small compact dust aggregates in low velocity collisions (Blum & Wurm 2008; Blum & Münch 1993)

The latter is in contrast to the growth of very fluffy or fractal aggregates which might create the large compact dusty bodies in the first place. In view of these results and considering recent other relevant collision experiments a straightforward scenario for planetesimal formation in collisions can be constructed. This is visualized in Fig. 11 and fits with coagulation simulations of a similar situation given in Johansen et al. (2008). In total, fractal growth, compaction, fragmentation, rebound and reaccretion might play together to lead to net growth of planetesimals even in the face of destructive collisions.

Acknowledgement

This work is funded by the Deutsche Forschungsgemeinschaft. We thank Georgi Paraskov for his initial contributions.